\documentclass[onecolumn]{svjour2}
\usepackage{graphicx}

\begin{document}

\title{Realism and single-quanta nonlocality}

\author{G.~S.~Paraoanu}
\institute{G.~S.~Paraoanu \at
Low Temperature Laboratory, Aalto University, P. O. Box 15100, FI-00076 AALTO, Finland \email{paraoanu@cc.hut.fi}
\at
Institute for Quantum Optics and Quantum Information (IQOQI),
Austrian Academy of Sciences, Boltzmanngasse 3, A-1090 Vienna, Austria}

\maketitle
\begin{abstract}
We show that local realism applied to states characterized by a single quanta equally and coherently shared between a number of qubits (so-called
W states) produces predictions incompatible with quantum theory. The origin of this incompatibility is shown to originate from the destructive interference of amplitude
probabilities associated with nonlocal states, a phenomenon that has no classical analog.
\PACS{03.65.Ud}
\end{abstract}

\section{Introduction}

%\section{Hidden variables and the Monty Hall problem}
%\label{MontyHall}

W states were first discussed in the literature in the early 1990's \cite{ZGH}, but they did not attract much attention until it was realized that under local operations and classical communication they generate an equivalence class with tripartite entanglement properties different from the GHZ states \cite{W}. Pure W states form a class of measure zero in the set of pure three-qubit states; however, if the state is mixed, then the measure is finite \cite{acin}.
Experimentally, three qubit W states have been realized with polarized photons created by parametric down-conversion, and the corresponding Bell inequality \cite{mermin} has been shown to be violated \cite{kiesel}.

But is it possible to show the incompatibility between local realism and quantum mechanics
in a more direct way, without using statistical correlations and Bell inequalities? A spectacular such example is of course the celebrated GHZ proof (also confirmed experimentally \cite{pan}), which uses perfect correlations and thus derives the local realistic prediction as an equality \cite{GHZ}. Unfortunately, such a proof cannot be extended to the case of W states, as shown by Cabello \cite{cabello1}. Another strategy, which relies on bringing up logical incompatibilities, was suggested by Hardy, who has first proposed an experimental test for the case of (two-partite) Bell states \cite{hardy}. Although the interpretation of the experimental proposal has sparked a controversy \cite{againhardy}, Hardy's setup has been eventually realized \cite{hardytest}, and his arguments have been generalized to  three-partite systems \cite{cabello2}, and very recently to N-partite systems \cite{arxiv}.

Despite the existence of such proofs, it is still unclear what is specific to W states - {\it i.e.} precisely which quantum effect is at the origin of this incompatibility, and if there is in the end any difference with respect to  the case of Bell or GHZ states.
In the same spirit of finding a logical contradiction between the predictions of  local realism and quantum mechanics rather than a mathematical relation (as in the EPR argument \cite{epr} and in \cite{hardy}), we show here that a surprisingly simple and clear argument can be obtained by analyzing measurement results which are predicted with a certain probability by a local realist approach but which are forbidden quantum-mechanically. In our argument, the origin of the contradiction between local realism and quantum mechanics is that the later allows destructive interference of amplitude probabilities of states defined nonlocally.
This proof does not apply to GHZ and to  Bell states, thus highlighting the fact that W states are in a class of their own from the point of view of their entanglement properties. This also sheds a new light on the discussion of what constitutes an entangled state versus a simple superposition for the case in which only one quanta is present, which again has sparked a significant debate \cite{enk}.

In this paper we use "local realism" in sense originating from the EPR paper \cite{epr}. By realism we mean the idea that the results of experiments are determined by intrinsic properties of particles which do not depend on measurement settings. The measurement just reveals these properties. By locality we mean the idea that the results of experiments cannot be influenced by events from which they are space-like separated (in the sense of the theory of relativity).

\section{Preliminary considerations}

W states are N-qubit states in which one excitation is equally and coherently distributed between the qubits. The problem of preparation of nonlocal single-photon states in quantum-optics setups has not been trivial \cite{againhardy}. Nowadays however, the preparation technique is much better understood, and W states - and even more complex quantum states - can be prepared in a reliable way with photons \cite{solved}, as well as with trapped ions \cite{ions}. In Appendix \ref{ap1} we present a generic way of preparing W states which could be suitable for the emerging field of superconducting circuit QED \cite{cqed}, where one can distribute a single quanta between three qubits and/or resonators and study the resulting three-party entanglement properties \cite{me}.

To set up the notations and the main assumptions for the rest of  the paper, let us consider first a qubit. A qubit is mathematically equivalent with a spin 1/2,  which can be oriented along $z$ either "up" $|\uparrow\rangle$ or "down" $|\downarrow >$  (the eigenvectors with eigenvalues $\pm 1$ of the Pauli matrix $\sigma_z$); in standard quantum-information notations these states are denoted by $|0\rangle$ and $|1\rangle$ respectively.

The result of a measurement along the $z$ direction performed on a spin prepared in the state $|1\rangle$ should be always 1. Similarly, if the spin is in the state $|0\rangle$ then we can predict with certainty that the result of a measurement along the $z$  direction will be -1.  A "realistic" interpretation of these results would simply take the parameters 0 and 1 in the kets $|0\rangle$ and $|1\rangle$ as representing "elements of reality". It would somehow assume that, whatever these numbers represent, they can be assigned to the qubit as pre-existent parameters. In some sense, we would like to endow these numbers with  the same degree of ontological consistency as a classical field.
In the words of Einstein, Podolsky, and Rosen \cite{epr}: "if, without in any way disturbing a system, we can predict with certainty [...] the value of a physical quantity, then there exists an element of physical reality corresponding to this physical quantity".

However, if now the qubit prepared in the state $|1\rangle$ is measured in the $xOy$ plane (transversal to the $Oz$ axis), then we get random results: $\pm 1$. Standard quantum-mechanical wisdom would teach us that these values are created by the very act of measurement and it is in principle impossible to predict the individual outcomes of such measurements. But a realist would disagree with this interpretation: equally well - and perhaps less mysteriously - it can happen that there is an additional "hidden" parameter which would determine the exact value of the spin along a transversal axis. The experimentalist does not have control over this parameter, and, therefore, in an ensemble of qubits prepared in the state $|1\rangle$, it will come randomly distributed. In fact, it is surprisingly simple to invent such a theory, which gives predictions in perfect agreement with quantum mechanics for measurements along any direction in space (see Appendix \ref{ap2}).
To distinguish local realist theories from quantum mechanics, in the following we will use a simple notation that accounts for the above intuition:
we denote  the local realistic state of a single qubit by an ordered array of parameters in a bracket $(\bullet ,\bullet ,\bullet , \bullet , ...)$, where the bullets represent the elements of reality to be assigned to each qubit (in our case there will be only two parameters, one referring to the value of the spin along the $z$ direction, the other along the $x$-direction). In the case of several qubits, due to the locality (separability) principle, we will assign elements of reality to each qubit separately; therefore an ensemble will be constituted by states of  the form
$[(\bullet ,\bullet ,\bullet , \bullet , ...); (\bullet ,\bullet ,\bullet , \bullet , ...); ....]$.

In the case of a $N$-qubit system however, it is not clear {\it a priori} how to assign such hidden variables to the qubits  -- should they be for example dependent on the hidden variable of the preparation qubit (see Appendix \ref{ap1}), and in which way? -- and why would in fact such a description be superior to assuming that certain values are created by the measurement itself? However, the EPR reality criterion offers a powerful answer to this. Suppose the qubits are spatially separated - such that actions performed at one site cannot influence causally the results at the other sites. If we can show that certain values emerge with probability 1 by performing local measurements on the other qubits, then we are guaranteed that these values are preexistent in the qubits, thus revealed and not created by measurement.  In the next section, we will apply the EPR reasoning to W states and show that this results in a logically inconsistent theory.

\section{Local realism and W states}

Consider a W state for three qubits,
\begin{equation}
|{\rm W}\rangle_{3} = \frac{1}{\sqrt{3}}\left(|100\rangle + |010\rangle + |001\rangle \right). \label{muru}
\end{equation}
We define the basis of the eigenvectors of the $\sigma_x$ Pauli operator by $\sigma_{x}|\pm\rangle = \pm |\pm\rangle$,
\begin{equation}
|+\rangle = \frac{1}{\sqrt{2}}\left( |1\rangle + |0\rangle\right),~~ |-\rangle = \frac{1}{\sqrt{2}}\left( |1\rangle - |0\rangle\right).
\end{equation}
Using this basis, we can rewrite Eq. (\ref{muru}) as
\begin{eqnarray}
|{\rm W}\rangle_{3} &=& \frac{1}{2\sqrt{3}}|1\rangle \left(|+\rangle|+\rangle
- |+\rangle|-\rangle - |-\rangle|+\rangle + |-\rangle|-\rangle \right)   \label{dec}\\
& & + \frac{1}{\sqrt{3}}|0\rangle
\left(|+\rangle|+\rangle - |-\rangle|-\rangle \right). \label{nov}
\end{eqnarray}
We now search for a local realistic theory: by analyzing possible measurement results, we aim at identifying the elements of reality needed to account for them. The first local realistic variable is the
value of  the spin along the z-direction. Indeed, this results from that fact that measuring $\sigma_z$ on any two qubits yields a fully predictable (with 100\% certainty) value for $\sigma_z$ on the third qubit. For example, suppose we have measured the values 1 and 0 on the first and second qubit; then, according to Eq. (\ref{muru}), we know that the results of a measurement of $\sigma_z$ on the third qubit will be 0. An ensemble of W states, in a local realistic description, will contain elements of the form $[(1, \bullet, \bullet , ...);(0, \bullet, \bullet , ...); (0, \bullet, \bullet , ...)]$ and permutations thereof. According to the EPR criteria of reality, this establishes the value of $\sigma_z$ as an element of  reality. Note also the restriction reflecting the single-quanta character of Eq. (\ref{muru}): there is always one 1 and two 0's in each element of  the statistical ensemble associated with the W states.

Next, we  claim that the values of a $\sigma_{x}$ measurement (the spin along the $x$- axis) must be also regarded as elements of reality. This follows by inspecting the line Eq. (\ref{nov}) in the expansion above for $|{\rm W}\rangle_{3}$. Indeed, suppose we have measured the value $0$ for $\sigma_z$ on the first site (which is an element of reality). Then, if we measure either of  the values $+$ or $-$ for $\sigma_x$ on the second site, we can predict with certainty (and without disturbing the system, since the measurements are local) an identical value $+$ (respectively $-$) for $\sigma_x$ on the third site. The same logic can be applied for any site, and this established the value of the spin in the $x$ -direction as an element of reality.

Thus, in a local realistic theory, each qubit will be described by $(\bullet, \bullet, ...)$. The first hidden variable will represent the value of  $\sigma_z$ with the restriction that, in a given element of the ensemble ({\it i.e.} for a 3-qubit triplet prepared in the $W$ state), there can be at most one 1 value for this parameter. The second bullet stands for the value of $\sigma_x$.

We now go back to examining the first line of the expansion of $|{\rm W}\rangle_{3}$ above (Eq. (\ref{dec})), according to which the following states (and, since the qubits are identical, all the rest of the permutations of single-qubit states inside the square brackets) will be present in such an ensemble
\begin{eqnarray}
&[(1, \bullet , ...);  (\bullet , +, ... );  (\bullet, + , ...)] ,\label{u}\\
&[(1, \bullet , ...);  (\bullet , +, ... );  (\bullet, - , ...)], \label{uu}\\
&[(1, \bullet , ...);  (\bullet , -, ... );  (\bullet, + , ...)], \label{uuu}\\
&[(1, \bullet , ...);  (\bullet , -, ... );  (\bullet, - , ...)] \label{uuuu}.
\end{eqnarray}
The probability of occurence of any of the states Eqs. (\ref{u}-\ref{uuuu}) is 1/12. We now use the restriction that an element of an ensemble can have only one value of $1$ for $\sigma_z$; since $\sigma_z$ is an element of reality, we infer that the values of the other spins are $0$. Note that this is a counterfactual reasoning, which is allowed by definition in the case of elements of  reality: we can claim that a parameter has a certain value even if we did not measure it but had measured another observable instead!
The elements of the ensemble Eq. (\ref{u}-\ref{uuuu}) can then be rewritten as
\begin{eqnarray}
&[(1, \bullet , ...);  (0 , +, ... );  (0, + , ...)] , \label{zero}\\
&[(1, \bullet , ...);  (0 , +, ... );  (0, - , ...)], \label{una} \\
&[(1, \bullet , ...);  (0 , -, ... );  (0, + , ...)],\label{doua} \\
&[(1, \bullet , ...);  (0 , -, ... );  (0, - , ...)]. \label{trei}
\end{eqnarray}
We now have to answer the question: which value should the spin that has $\sigma_z =1$ assume as element of reality for the x-component of the spin? Obviously, it can be either $+$ or $-$. Suppose it  is $+$; then for example Eq. (\ref{una}) becomes $[(1, +, ...);  (0 , +, ... );  (0, - , ...)]$. But this means that it is possible to have a situation in which we measure $\sigma_z$ on the second qubit (with the result 0) and $+$ respectively $-$ as  results for $\sigma_x$ measurements on the first and third qubit. However, this is forbidden quantum-mechanically (and we accept that quantum mechanics agrees with the experiment): indeed $|W\rangle_{3}$ has null component on $|+\rangle|0\rangle |-\rangle$,  $_{3}\langle W |+ 0 -\rangle =0$. We are then forced to assign the value $-$, and Eq. (\ref{una}) becomes $[(1, -, ...);  (0 , +, ... );  (0, - , ...)]$. But again, this means that by choosing to measure $\sigma_x$ on the first and second qubit and $\sigma_z$ on the third, with the results $-$, $+$, and $0$ respectively. However, the component of $|W\rangle_{3}$ on the state $|-\rangle |+\rangle |0\rangle$ is again zero, $_{3}\langle W |- + 0\rangle =0$, so again these states cannot exist in the ensemble. We have then obtained a contradiction. A similar logical incompatibility is obtained for states of the type Eq. (\ref{doua}). Since the states Eqs. (\ref{una}, \ref{doua}) appear each 1/12 times in a measurement over a large ensemble of 3-qubits prepared in the state $|W\rangle_{3}$, we conclude  that by this argument local realism is shown inconsistent in 1/6 cases.
Note also that it is possible to start with the states Eqs. (\ref{zero}, \ref{trei}) but then the
argument proceeds along a slightly different line (see Appendix \ref{ap3}).

The extension of this argument to  $N$-qubits W states with $N$ an odd integer number is immediate: a $|{\rm W}\rangle_{N}$ state is defined, in the $\sigma_z$ basis of each site, as
\begin{equation}
|{\rm W}\rangle_{N} = \frac{1}{\sqrt{N}}\left(|1000 ..0\rangle + |0100 ..0\rangle + |0010 ..0\rangle + .... |0000 ...1\rangle \right) .\label{alll}
\end{equation}
We first establish the values of the "spin" along the $z$ and $x$ directions as elements of reality for each site, by the same procedure as before.
Then we proceed by analyzing what happens when we measure the value $1$ for $\sigma_z$, say on the first site, and then have a string of $+$ and $-$ results for measurements along $\sigma_x$ for the remaining $N-1$ sites. Suppose that the result of the measurement is such that there are equal numbers of $+$ and $-$ results. Then the local realistic description of such an occurrence is
\begin{equation}
[(1, \bullet, ... ); (\bullet , +, ... ); (\bullet , - , ...); (\bullet , + , ....), ... , (\bullet , - , ....)]  .
\end{equation}
Now, since the value of $\sigma_z$ is also an element of reality, it must then have been the case that $\sigma_z$ on  the sites $2,3, ..., N$ was zero, therefore we have  \begin{equation}
[(1, \bullet, ... ); (0, +, ... ); (0 , - , ...); (0, + , ....), ... , ( 0, - , ....)].
\end{equation}
Which value of $\sigma_x$ shall we associate with the first site? It does not matter: the important fact is that it is possible to ascribe one. Suppose that it is $+$. Then the local-realistic description of this particular member of the ensemble is $[(1, +, ... ); (0, +, ... ); (0 , - , ...); (0, + , ....), ... , ( 0, - , ....)]$.
We thus obtain the local realistic prediction that sometimes we will get for example the result $+,0, -, ..., +, -$. However, this is strictly forbidden by quantum mechanics: the probability to get such  a result, as can be checked using the state  Eq. (\ref{alll}), is exactly zero.

\section{Comments and conclusion}

What is the cause of the failure of the local-realist description? Let us look in detail at what happens when we have three sites and we measure the last two along the $x$ direction , getting for example the result $+$ for the second site and $-$ for the third. What do we expect for the first site? Let us analyze the three terms in the superposition Eq. (\ref{muru}). Under projection onto  $|+\rangle$ for the second site and $|-\rangle$ for the third, $|100\rangle$ becomes $-|1+-\rangle$, the term $|010\rangle$ becomes $-|0+-\rangle$, and $|001\rangle$ becomes $|0+-\rangle$. There is destructive interference between the last two terms, leading to zero probability of getting the result $0$ under a $\sigma_z$ measurement on the first site. But what has interfered here? After all, the sites and measurements can be spatially separated, so there is no way even for a "hidden" wave emerging after one measurement on one site to reach any of the others. Interference of probability amplitudes is indeed a mind-boggling consequence of quantum physics, with no classical equivalent.

It is also interesting to note that one needs at least 3 spins to violate the local realistic assumption.
Consider for example the case of Bell state $|{\rm W}\rangle_{2}$,
\begin{equation}
|{\rm W}\rangle_{2} = \frac{1}{\sqrt{2}}\left(|1\rangle|0\rangle + |0\rangle|1\rangle \right) =
\frac{1}{\sqrt{2}}\left(|+\rangle |+\rangle -|-\rangle|-\rangle \right). \label{scunto}
\end{equation}
A local realistic theory would then consider an ensemble consisting of correlated
pairs such as $|1+\rangle , |0 +\rangle$, $|1-\rangle , |0-\rangle$, distributed in the ensemble with equal weights. There is no contradiction one can obtain by attempting an argument of the form above. The only claim one can make is about incompleteness, which is precisely the message of the original EPR argument.
But with a ${\rm W}$ state, when rotating each qubit to the $x$ direction, the structure of the expansion is not preserved, as in the
case of Bell states (see Eq. (\ref{scunto}) as compared to Eq. (\ref{muru}, \ref{dec}, \ref{nov})).
Also, the argument does not work for GHZ states, which is perhaps surprising since these states have maximal correlations. Indeed, consider
\begin{equation}
|{\rm GHZ}\rangle_{N} = \frac{1}{2}\left(|11 ...1\rangle + |00 ...0\rangle \right) .
\end{equation}
What are the chances of getting a spin up, according to classical realism? Obviously 1/2.
Making now $N-1$ measurements along $x$ would collapse the wavefunction onto a state $\frac{1}{2}\left(|1\rangle \pm |0\rangle\right)$, so there is a probability 0.5 to get the results 0 or 1 when measuring the remaining qubit along $z$. Curiously, precisely the maximal correlations which are the hallmark of these states preclude the interference between amplitude probabilities.

Finally, how can it be that we have obtained an effect which supposedly is related to entanglement from
seemingly just a superposition of a single quanta (say a photon) amongst several sites?
The underlying reason for this, which connects also to the discussion of entanglement versus superposition \cite{enk}, is that vacuum might be an empty state, but the fact that it contains no particle is a well-defined piece of information. The amplitude probabilities associated with this information behaves like any other piece of quantum-mechanical information - for example they interfere, {\it etc.}. In general, our argument supports the idea that quantum physics can be understood as a theory about limited amounts of information \cite{brukner}. In this reconstruction of quantum mechanics, an elementary system can carry only one bit of information. In the case of a W state, this bit of information is derived from the preparation procedure: simply put, the preparation bit (see Appendix \ref{ap2}) contains one bit of information; the time it interacts with the other qubits is set so that we are sure that this bit  is completely transferred to the $N$ qubits (as can be checked by repeating the  experiment enough many times and measuring  the $\sigma_z$ of the probe qubit).
Then we end up with exactly one bit of (classical) information spread across $N$ locations. On the other hand, the logic of local realism forces us to associate additional bits of information to each site. As we have shown, this results in a contradiction with quantum-mechanical predictions.

In conclusion, we have given a refutation of  local realism without inequalities. We have shown that local realism predicts the occurrence of certain results, while quantum mechanics predicts that the probability of obtaining such results is exactly zero.

\section{Acknowledgements}
%\acknowledgements

 The research for this work was done under a John Templeton Fellowship, which allowed the author to spend the summer of 2009 at IQOQI Vienna. Special thanks go to my
hosts in Vienna, Prof. A. Zeilinger and Prof. M. Aspelmeyer, who have
made this visit possible, and additionally to the scientists in the institute for many enlightening discussions. Also, support from the Academy of Finland is acknowledged (Acad. Res. Fellowship 00857, and projects 129896, 118122, and 135135).

\appendix

\section{A realistic model for a single qubit}
\label{ap2}

A hidden-variable model for spin-1/2 has been discussed already by Bell and refined by Mermin \cite{bell-mermin}. Below we give a brief review of this model. Although our analysis for W states is independent on any particular hidden-variable interpretation, the Bell-Mermin model is very useful for having a mental picture of how a realistic description of a qubit would look like.

 Any state of a spin 1/2 can be regarded as the eigenstate of the spin operator along a direction $\vec{n}$, that is, $\sigma_{\vec{n}}|\uparrow\rangle_{\vec{n}}=|\uparrow\rangle_{\vec{n}}$. Suppose that we have prepared the system in such a state.
 The general form of observable $A$ in the spin-1/2 space is  $A=a_{0} + \vec{a}_{1}\vec{\sigma}$, and the values taken by such observables when measured, are $a_{0}\pm a_{1}$, where $a_{1}$ is the magnitude of the vector $\vec{a}_{1}$. The statistics of these measurements on a given state $\vec{n}$ is such that $_{\vec{n}}\langle\uparrow|A|\uparrow\rangle_{\vec{n}} = a_{0} + \vec{a}_{1}\vec{n}$.
A hidden variable theory would postulate a unit vector  $\vec{m}$, which is not controllable by the experimentalist, and which however fully determines the result  $v_{\vec{n}}(A)$ of $A$ along $\vec{n}$ by the "measurement theory"
\begin{equation}
v_{\vec{n}}(A) = \left\{\begin{array}{cc} a_{0} + \vec{a}_{1} & {\rm if}  ~(\vec{m}+\vec{n})\vec{a}_{1}>0 ,\\ a_{0} - \vec{a}_{1} & {\rm if} ~(\vec{m}+\vec{n})\vec{a}_{1}<0 . \end{array}\right.
\end{equation}
It is then straightforward to show that, by averaging over the results of any measurement of the observable A (over the solid angle $\Omega_{\vec{m}}$), one recovers indeed the quantum-mechanical predictions
\begin{equation}
\int \frac{d\Omega_{\vec{m}}}{4\pi}v_{\vec{n}}(A)  = \frac{1}{2}\int_{-1}^{1}d(\cos \theta )\left\{\begin{array}{cc} a_{0}+a_{1} &{\rm if} \cos\theta >-\cos\xi \\ a_{0}-a_{1} &{\rm if} \cos\theta <-\cos\xi \end{array}\right. = a_{0}+\vec{a}_{1}\vec{n},
\end{equation}
where $\xi$ is the angle between $\vec{a}_{1}$ and $\vec{n}$. In the particular case of a spin polarized along the $z$ direction and measured along the $x$ direction, we readily find out that the results are randomly distributed.

\section{Preparation of W states}
\label{ap1}

We consider a system of $N$ qubits (or 1/2 spins) which are all in the ground state (or polarized in the "down" direction). The qubits are assumed to be noninteracting. We now bring in the "preparation" qubit (indexed with "p"), which is excited in the state $|1\rangle$, and have it interact with them by the standard rotating-wave approximation Hamiltonian
\begin{equation}
H= -t \sum_{j=1}^{N}\left(\sigma_{p}^{+}\sigma_{j}^{-}+\sigma_{p}^{-}\sigma_{j}^{+}\right).
\end{equation}
It is easy to understand the dynamics of this Hamiltonian if we introduce the notation
\begin{equation}
S^{\pm} = \frac{1}{\sqrt{N}}\sum_{j}\sigma_{j}^{\pm}.
\end{equation}
The algebra on the restricted subspace with just one spin flipped up is then very simple: we have
\begin{equation}
S^{+}|00... 0\rangle = |{\rm W}\rangle_{N} ,
\end{equation}
and
\begin{equation}
S^{-}|{\rm W}\rangle_{N} = |00 ... 0\rangle .
\end{equation}
Using  this collective operator, the Hamiltonian becomes
\begin{equation}
H = -t\sqrt{N}(\sigma_{p}^{+}S^{-} + \sigma_{p}^{-}S^{+}),
\end{equation}
which describes Rabi oscillations between the two states. To create a W state is therefore enough to make a $\pi /2$ rotation in this two-dimensional subspace,
that is, to act with the Hamiltonian for a time $\tau_{\pi /2} =\pi /2t\sqrt{N}$.

\section{A different version of the proof}
\label{ap3}

Why is it not possible to construct a local realistic theory which for example would simply not contain the elements forbidden by quantum mechanics? Suppose we list all possible combinations and eliminate those which do not satisfy the two rules we have discovered: the first, that in any element of the statistical ensemble there are always two 0's and one 1, and the second, that in any such element the combination of a $0$ with a $+$ and a $-$ is excluded. By using these two rules, out of $2^{6}=64$ possibilities, we are left only with 6 states, namely:
\begin{eqnarray}
&&[(1,+), (0, +), (0, +)] ,\label{mmm}\\
&&[(1,-), (0,-), (0, -)]  ,\label{mmmm}
\end{eqnarray}
and permutations. However, according to Eq. (\ref{dec}) it is possible also to have results such as $1, +, -$ {\it etc.}; but these states do not occur in the ensemble described by  Eq. (\ref{mmm}, \ref{mmmm}). Another way of  obtaining a contradiction is to notice that, by way of Eq. (\ref{dec}, \ref{nov}),  we have to assign equal
equal occurrence probabilities (equal to 1/12) to the states  Eq. (\ref{mmm}, \ref{mmmm}): but since are only 6 states the sum of probabilities for the ensemble would be 1/2 instead of 1.

Finally, we notice that in  this paper we have derived a contradiction with local realism from only two sets of measurements: one of $\sigma_z$ on all qubits and the other one of $\sigma_z$ on one qubit and $\sigma_x$ on the rest. Our argument is more economical than previous ones \cite{cabello2}, which obtain contradictions with local realism by involving a third set of measurements, namely of $\sigma_x$ on all qubits. Indeed,
by inspecting Eqs. (\ref{mmm},\ref{mmmm}), one sees that local realism predicts that in such measurements all the results will be equal, which is obviously contradicted by the quantum-mechanical predictions based on Eq. (\ref{muru}). However, it vas very recently noticed \cite{arxiv} that involving a third set of measurements
has a beautifully surprising advantage for $N$-sites W states. In this case, if $N$ is large, the probability of getting the result predicted by local realism (all $\sigma_x$ measurements give the same result) becomes very small: most of the time, quantum mechanics will win!

\end{document}